**ONSHORE-WIND ENERGY | RENEWABLE ENERGY**

# Wind energy potential of Germany
## Limits and consequences of large-scale wind energy use

Axel Kleidon

*The transition of our energy system to renewable energies is necessary in order not to heat up the climate any further and to achieve climate neutrality. The use of wind energy plays an important role in this transition in Germany. But how much wind energy can be used and what are the possible consequences for the atmosphere if more and more wind energy is used?*

The German government wants to achieve the goal of a climate-neutral energy system by 2050. This goal envisages a strong expansion of wind energy, and 2% of Germany's surface area is to be made available for this purpose. Scenarios from various institutions translate this into about 150 - 200 gigawatts of installed capacity, contributing 330 - 770 TWh per year to electricity generation. As an example, we refer to the studies by Agora Energiewende and the German Wind Energy Association [1, 2]. Currently, only 56 GW of capacity is installed, distributed across 28230 wind turbines located in Germany at the end of 2021 [3]. These wind turbines generated 90.3 TWh/year of electricity, contributing just under 16% to the current electricity generation of 570 TWh per year (as of 2021, [4]). This means that we need a strong increase in wind turbine installations in the coming decades to achieve the goal of climate neutrality.

But how much wind energy is there in Germany, and how much of it can be used? What are the effects on the atmosphere if more and more kinetic energy is extracted from it by the wind turbines? While such energy scenarios often focus on what is technically possible, here we want to look at the physics of the atmosphere and derive simple estimates that can provide answers to these questions.

## How wind energy comes to Germany

In order to estimate how much wind energy can be used in Germany, we first look at where wind energy comes from and how much of it comes to Germany. Wind turbines in Germany mainly use large-scale winds associated with the high and low pressure systems in the mid-latitudes. These are directly linked to the large-scale atmospheric circulation. This circulation is driven by planetary differences in solar radiation: tropical areas absorb more solar radiation than mid-latitudes and polar areas, thus the tropics are warmer and the poles colder. Temperature differences lead to different air densities, these cause air pressure to drop less with altitude in warm areas, which generates potential energy. This in turn is associated with differences in air pressure in the middle atmosphere, where most of the weather activity takes place. Air is accelerated, mass is moved and rearranged, thus potential energy is dissipated, heat is transported and the differences in solar heating are depleted. Kinetic energy plays a central role in this process, as it is directly linked to motion and heat transport. It is embedded as a form of energy in the conversion chain from incident solar energy to heating differences, which leads to potential energy from which kinetic energy is extracted, which is ultimately converted back into heat by friction and radiated from the Earth into space in the form of longwave radiation.

**FIG. 1: WIND AND DISSIPATION**

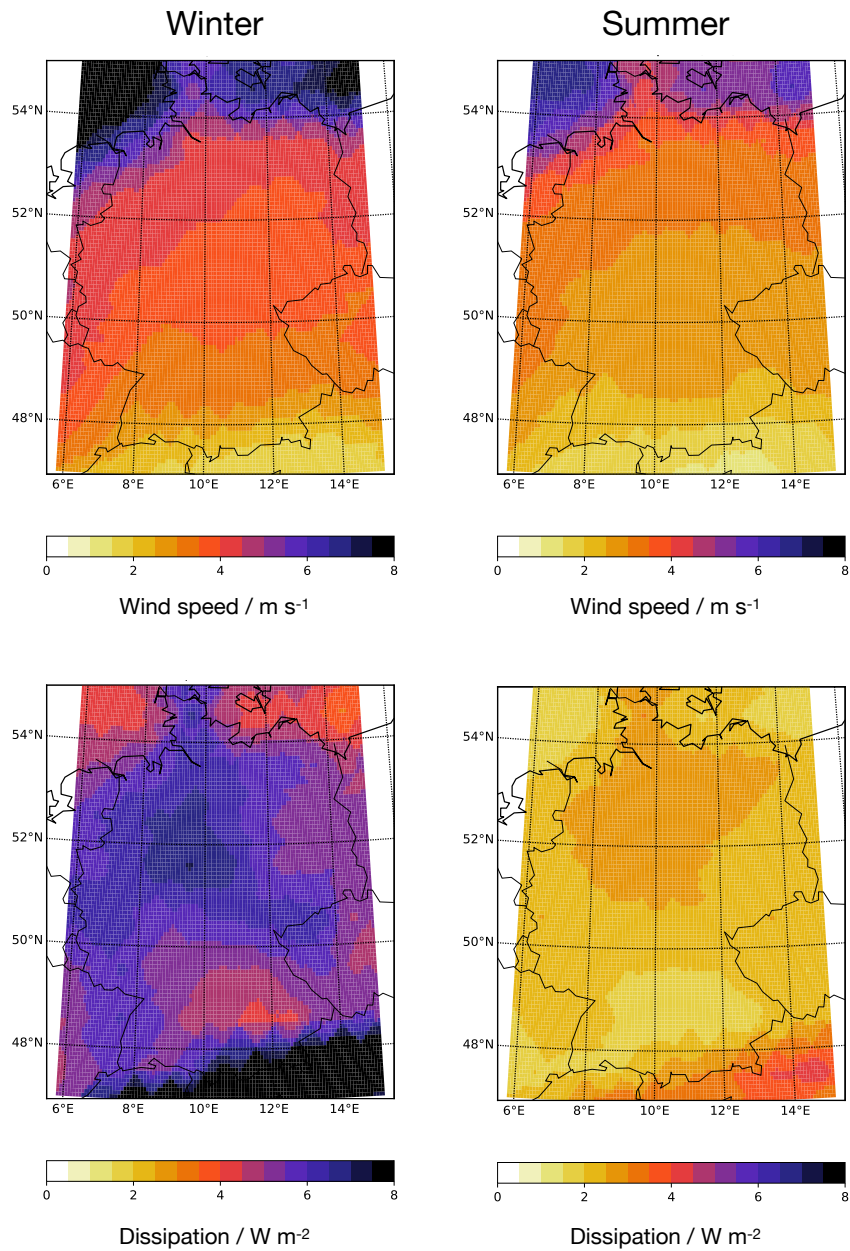

*Wind speeds at 10m height and dissipation for winter (December-February) and summer (June-August). The maps show long-term means over the years 1980-2010 of the weather reanalysis dataset ERA-5 [7].*

The power contained in the large-scale circulation that generates kinetic energy can be easily estimated with the help of thermodynamics. To do this, we consider the generation of air movement as the result of a heat engine (or "power plant", [5]) driven by the differential heating by solar radiation and the associated temperature differences. The transported heat represents the heat flux that drives the heat engine, together with the temperature difference. It is thus relatively easy to estimate that the large-scale circulation only provides about 2 W m$^{-2}$, i.e. generates kinetic energy from heating differences [6]. This order of magnitude agrees very well with the power estimated from observations, so the atmosphere works as hard as it can.

The generated motion is then dissipated by friction, predominantly in the so-called boundary layer near the Earth's surface, i.e. it is converted into heat. Because of the conservation of angular momentum, however, the friction is not evenly distributed, but is concentrated in the mid latitudes, precisely where the westerly winds blow relatively strongly and the high and low pressure systems alternate, in a part of the Earth where Germany is also located. This means that the frictional loss over Germany is about twice as high as the global average, at about 4 W m$^{-2}$ (Figure 1). In Figure 1 we can also see the seasonal variations very clearly: In winter, the frictional losses are much higher than in summer. This, in turn, can be explained by the atmospheric "power plant": in winter, the differences in radiation between the tropics and the polar regions are greater, the power plant works harder, and consequently there are stronger winds.

For Germany we can summarize this as follows: Wind energy is predominantly generated in the so-called free atmosphere, i.e. far from the surface, from where it is brought to the surface and is dissipated by friction. This means that near-surface wind energy use can have practically no effect on the large-scale circulation. After all, wind turbines only use the kinetic energy that is lost to surface friction anyway, while weather patterns take their course far above, decoupled from the surface. This has a greater, and more direct, consequence for the supply of wind energy if it is to be used by many turbines over large areas. This supply comes from above, and at a relatively low rate of 4 W m$^{-2}$. With this rate, we can already estimate how much wind energy is naturally lost over Germany due to friction. To do this, we take the 4 W m$^{-2}$ and multiply it by Germany's surface area of 357 000 km$^2$, and thus obtain about 1430 GW, which corresponds to 12 500 TWh or 45 100 PJ per year. For comparison, the average primary energy consumption of Germany during 2021 was 393 GW, or 12 400 PJ per year [4], which is about 27% of the natural friction. From this comparison we can see that a strong expansion of wind energy could well use a considerable part of the energy that the atmosphere naturally brings to Germany and loses there through friction.

## Large-scale limits to wind energy use

To estimate how much of the input of wind energy into the boundary layer can be used by wind turbines, we follow how the kinetic energy from the free atmosphere reaches the surface (Figure 2). The kinetic energy is transported to the Earth's surface by the downward flow of horizontal momentum. Our starting point for the description is therefore the momentum balance, since momentum is conserved. Kinetic energy, on the other hand, is not conserved. When momentum is transported from the free atmosphere to lower levels at lower velocities, kinetic energy is lost as it is dissipated into heat. At the lower end of the transport, the remaining energy is lost by friction at the Earth's surface. In the absence of wind turbines, this balance determines the wind speed near the surface that reflects both, the driver from above and the frictional properties of the surface. This can be seen nicely on the maps in Figure 1: Sea surfaces are generally smoother, so surface friction occurs at higher wind speeds than over land, although kinetic energy is lost in the boundary layer at overall about the same rate.

When more and more wind turbines are introduced into a region, then the balance of momentum and friction is changed because the wind turbines extract part of the kinetic energy (Figure 3). This leads to another term in the momentum balance. Instead of only surface friction, a part of the momentum is brought into the ground via the wind turbines. The wind speed thus decreases in the region. This then shifts the balance of how the kinetic energy is dissipated or converted. With more wind turbines, the loss due to surface friction is reduced, but the lower wind speeds increase the loss due to mixing in the boundary layer. This then sets a limit on how much energy wind turbines can extract from the boundary layer. Since the energy production of wind turbines

**FIG. 2: KINETIC ENERGY NEAR THE SURFACE**

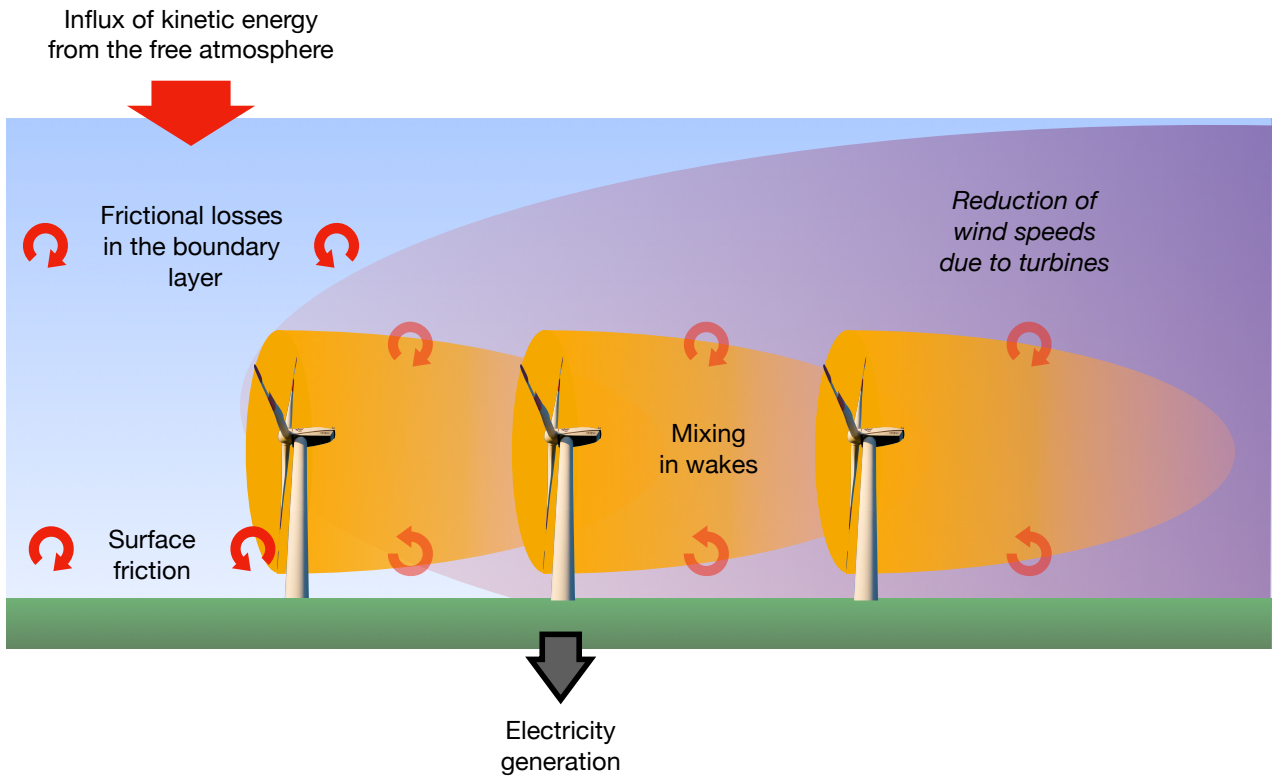

***Kinetic energy is brought down from the free atmosphere to the surface. During the process, it is converted into heat by mixing in the boundary layer and surface friction, or it is converted into electrical energy by wind turbines. Some of it is then lost through mixing in the wake of the turbines.***

increases with the third power of wind speed, each individual wind turbine thus produces less electricity, so they become less efficient.

A simple formulation of the momentum balance shows that the turbine efficiency (also called capacity factor) decreases approximately linearly with the rate at which kinetic energy is removed from the atmosphere by the turbines (see box: "Vertical Kinetic Energy transport"). The efficiency of the turbine describes how much electricity the turbine produces in relation to how much it could produce, i.e. the capacity of its generator. The efficiency thus corresponds to the utilisation of the turbine. A single turbine has the highest efficiency, which we call $f_{cap,max}$. It can be determined from current wind turbines and is about $f_{cap,max} \approx 20\%$ in Germany [8]. With more and more wind turbines in a region, the efficiency ($f_{cap}$) decreases because of the reduced wind speeds:

$$f_{cap} \approx f_{cap,max} (1 - \beta \times (f_{cap,max} \times Y_{ic})/Y_{max})$$

The effect of the weakened winds is described here by the expression in the brackets. The parameter $\beta$ determines how strong the decrease in efficiency is with the addition of wind turbines. Climate model simulations yield a value of $\beta \approx 0.8$, from simple theory the value is somewhat lower with $\beta = 1/\sqrt{3} \approx 0.58$ (see box). The number of wind turbines enters this expression by the total installed capacity distributed over a region with a certain area ($Y_{ic}$, with the unit MW km$^{-2}$ or W m$^{-2}$ ). The input of kinetic energy through the atmosphere enters via $Y_{max}$,

# VKE: Vertical Kinetic Energy transport

To estimate yield reductions in regional wind energy development, we formulate a momentum balance of the lower atmosphere, i.e. the part of the atmosphere where, in the absence of wind turbines, the kinetic energy of the atmosphere is dissipated by friction and converted into heat [6]. We use the momentum balance to describe the reduction in wind speed of the lower atmosphere as a function of the number of turbines installed. This then allows us to determine how the input of kinetic energy from the atmosphere is split between losses due to mixing in the boundary layer and surface friction, and energy production by wind turbines (Figure 2). From this we derive an approximation that determines how the average efficiency of wind turbines decreases when wind energy is expanded regionally.

## Momentum balance

Conservation of momentum applies to the transport of kinetic energy from the free atmosphere to the surface. The momentum introduced is balanced with the frictional force at the surface, $F_{Friction}$, and the momentum extraction by the wind turbines, $F_{Turbines}$ (all described per unit area):

$$F_{in} = F_{Friction} + F_{Turbines}$$

In climate research, surface friction is typically described by turbulent friction:

$$F_{Friction} = \rho\, C_d\, v^2$$

where $\rho$ is the air density (about 1.1 kg m$^{-3}$), $C_d$ is the drag coefficient, and $v$ is the wind speed of the lower atmosphere. The drag coefficient depends on the roughness of the surface, the stability of the air, and the reference height. Typical values for so-called neutral conditions at 100m are 0.001 for the sea, 0.005 for grass, and 0.01 for forest.

Momentum extraction by the wind turbines is described in a similar way. The drag coefficient is replaced here by the cross-sectional area, $A_{Rotor}$, spanned by the rotor blades of the turbines:

$$F_{Turbines} = n\, \rho/2\, \eta\, A_{Rotor}\, v^2$$

Here, the turbine density $n$ is the number of wind turbines $N$ per area $A_{Region}$ (i.e. $n = N/A_{Region}$), and the power coefficient $\eta$ enter here. For modern wind turbines that do not operate at their capacity, this coefficient is typically about $\eta \approx 0.42$ [8].

If we combine the equations, we can express wind speed as a function of momentum input, $F_{in}$, and installed wind turbines:

$$v = (F_{in}/(\rho\, C_d + n\, \rho/2\, \eta\, A_{Rotor}))^{1/2} = f_{Red}^{1/3} \times v_0$$

Without wind turbines, the wind speed assumes a value of $v_0 = (F_{in}/\rho\, C_d)^{1/2}$. Thus, the reduction effect due to the wind turbines can be described by the factor $f_{Red}$:

$$f_{Red} = (1 + n\, \eta\, A_{rotor}/(2\, C_d))^{-3/2}$$

From this equation we can already see that more wind turbines (higher $n$) must lead to a lower $f_{Red}$ and reduced wind speeds (lower $v$).

## Kinetic energy balance

The balance of kinetic energy then looks like this: The input of kinetic energy, $J_{ke}$, associated with the vertical transport of momentum is described by the momentum flux as well as the velocity of the free atmosphere, $v_{in}$:

$$J_{ke} = F_{in}\, v_{in}$$

This input corresponds to the rate of energy lost through friction under natural conditions, i.e. dissipation, as shown in Figure 1. It is balanced with the various terms that dissipate kinetic energy: the energy output by the wind turbines $Y$, the surface friction, $D_{Friction}$, and the frictional losses due to mixing within the boundary layer and in the wake of the turbines, $D_{Boundarylayer}$:

$$J_{ke} = Y + D_{Friction} + D_{Boundarylayer}$$

The energy yield of the wind turbines is described by

$$Y = F_{Turbines}\, v = f_{red}\, n\, \rho/2\, \eta\, A_{Rotor} \times v_0^3$$

the loss due to surface friction by

$$D_{Friction} = F_{Friction}\, v = f_{Red}\, \rho\, C_d \times v_0^3$$

and by mixing by

$$\begin{aligned} D_{Boundarylayer} &= F_{in}\, (v_{in} - v) \\ &= J_{ke} - f_{Red}\, (2\, C_d + n\, \eta\, A_{Rotor}) \times \rho/2\, v_0^3 \end{aligned}$$

The distribution of these terms and how these change with more turbines is shown in Figure 3.

## Yield and efficiency of the wind turbines

The total yield from the wind turbines increases with the number of wind turbines (higher $n$). For very small values of $n$, this expression gives approximately

$$Y \approx n\, \rho/2\, \eta\, A_{Rotor}\, v_0^3$$

With increasing $n$, the power generation increases, but because the wind speed and $f_{red}$ decrease with $n$, the result is a maximum of generated energy

$$Y_{Max} = 2/3^{3/2} \times \rho\, C_d\, v_0^3 \approx 38\%\, D_{Friction}(n = 0)$$

for $n_{opt} = 4\, C_d/(\eta\, A_{Rotor})$ and a value of $f_{Red} = 2/3^{3/2}$. Thus, only a maximum of about 38% of the energy dissipated by surface friction under natural conditions without wind turbines can be used.

We obtain the average efficiency of the turbines, $f_{cap}$, by dividing the yield $Y$ by the installed capacity density, $Y_{ic}$, i.e. the installed capacity of turbines over an area in units of MW km$^{-2}$. This can then be described by the following equation:

$$\begin{aligned} f_{cap} &= f_{cap,max} \times (1 - f_{red} \times Y/Y_{max}) \\ &\approx f_{cap,max} \times (1 - 0.58 \times Y/Y_{max}) \end{aligned}$$

The reduction in efficiency thus depends primarily on the energy extraction by the turbines, and is largely independent of the details of the turbines. This expression is used in the main text to estimate the reduction effect for different expansion scenarios.

### FIG. 3: EFFECTS OF WIND ENERGY USE

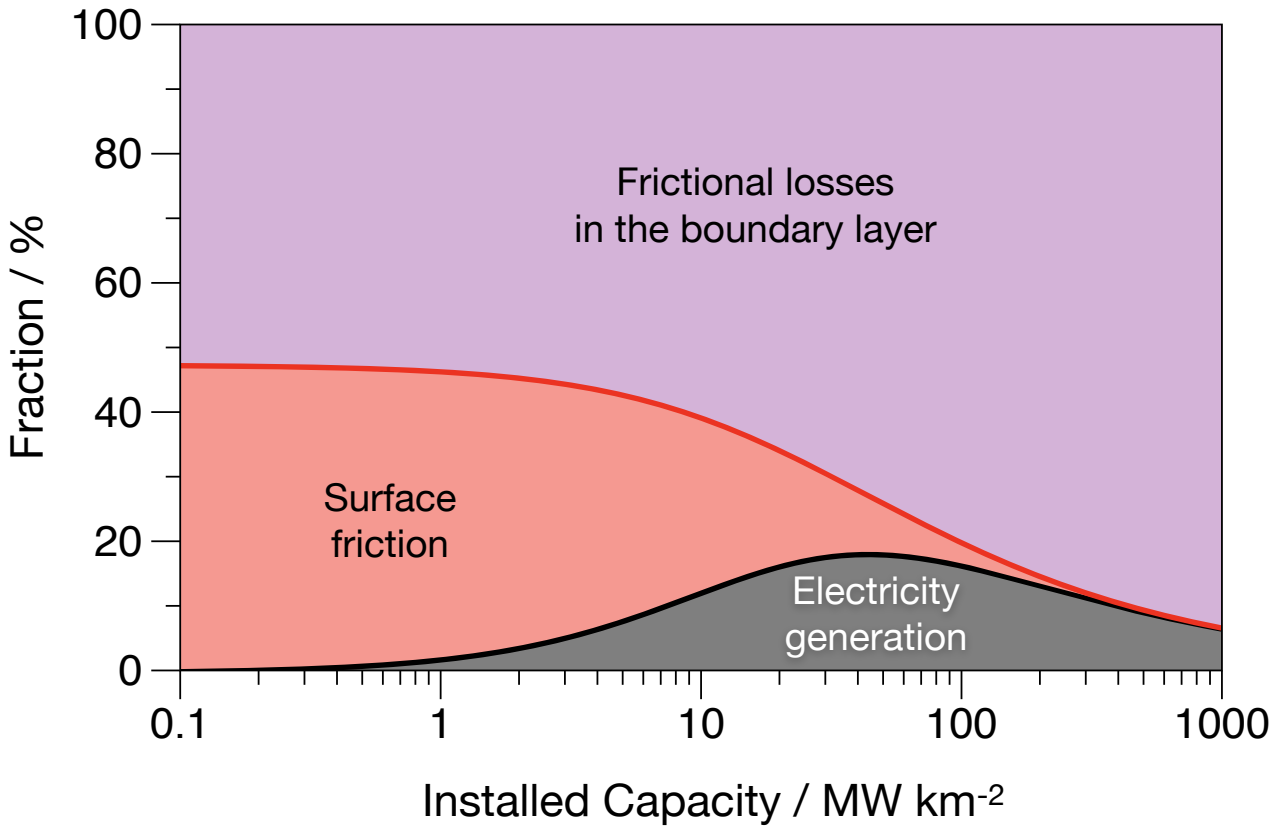

*The more wind turbines extract energy from the atmosphere (grey), the more of the kinetic energy input is lost through mixing above the turbines in the boundary layer (purple). This leads to a relatively low limit on how much wind energy can be used by wind turbines.*

which describes the maximum usable part for wind energy generation (see Box).

Let us briefly illustrate the plausibility of the equation before we apply it. If the number of turbines installed in a region is very small, $f_{cap,max} \times Y_{ic}$ is much smaller than $Y_{max}$, and the expression in brackets in the equation is about 1. Then the efficiency of the wind turbines is close to the maximum value of an isolated wind turbine, i.e. $f_{cap} = f_{cap,max}$. More turbines lead to more extraction of wind energy from the atmosphere, so $f_{cap,max} \times Y_{ic}$ becomes larger, and the expression in brackets is smaller than 1. The efficiency of the wind turbines thus decreases. This effect is stronger in regions and seasons where the atmosphere contributes little kinetic energy, so the rate $Y_{max}$ is comparatively low. If more wind turbines are installed in the region than the maximum possible yield, i.e. $f_{cap,max} \times Y_{ic} > Y_{max}$, then this linear approximation no longer applies and the yield must be determined differently. However, these are unrealistic scenarios of a drastic expansion of wind energy.

### Estimates for Germany

With the equation, we can now consider different expansion scenarios that distribute certain installed capacity over certain areas, and thus estimate the effects of the energy extraction by the

## FIG. 4: ELECTRICITY YIELDS IN GERMAN STATES

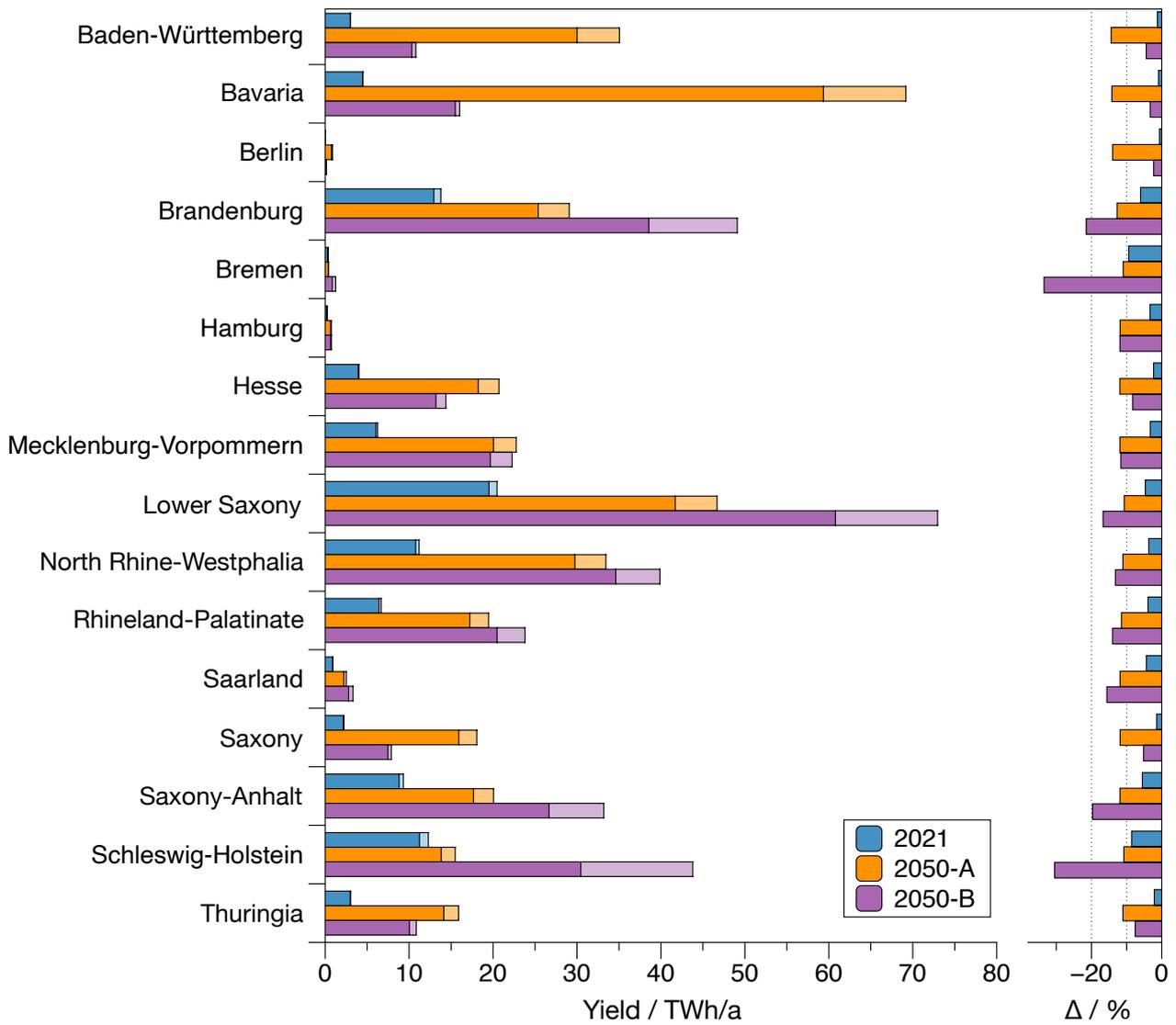

*Electricity yields from wind energy estimated for the current installed capacity and for two expansion scenarios for 2050 by state. The light-coloured bars show the estimates that do not take the effect on the atmosphere into account, the dark-coloured bars include the reduction effect by the wind turbines on wind speeds. The relative reductions compared to the case without reduced wind speeds are shown on the right. For comparison: Germany's total electricity demand in 2021 was 570 TWh.*

wind turbines on the electricity yield and on the atmosphere. These scenarios determine the value of $Y_{ic}$.

We first use this for the wind turbines already installed in 2021. To do this, we divide the 56 GW of installed capacity by the area of Germany of 357 000 km² and obtain an installed capacity density of $Y_{ic}$ = 56 x 10⁹ W / (357 x 10⁹ m²) = 0.16 W m⁻². We estimate the maximum possible yield $Y_{max}$ from the mean value of about 4 W m⁻². About half of this is lost through mixing in the boundary layer, and a maximum of 38% of the rest can be used (see Box). We thus obtain $Y_{max}$ = 0.76 W m⁻². With the equation we get a value of $f_{cap}$ = 19.3% for the efficiency instead of 20%, which is about 3.5% lower. We can now multiply this value by the installed capacity of 56 GW and get an annual electricity yield of 95 TWh instead of 98 TWh without the wind reduction effect. Both estimates compare quite well to the reported contribution of wind energy of 90 TWh per year [3]. So the effect of wind turbines on wind speeds is currently quite small. A closer look, which includes the distribution of the 56 GW across the federal states, is shown in Table 1 and Figure 4, and leads to very similar results overall (yield of 94 TWh per year, i.e. a reduction of 4.4%).

We can now take a scenario of 200 GW of installed capacity and repeat this estimate. This results in a capacity density of $Y_{ic}$ = 200 x 10$^9$ W/ (357 x 10$^9$ m$^2$) = 0.6 W m$^{-2}$, and the efficiency drops by about 13% to $f_{cap}$ = 17.5%. The annual electricity yield decreases accordingly, and instead of 350 TWh without the reduction effect, it still delivers 307 TWh per year. Figure 4 shows the more precise estimate on the scale of states as "Scenario 2050A"; here the yield drops to 313 TWh per year, i.e. about 11% less. Also shown is a "Scenario 2050B", in which it is assumed that wind energy use will be expanded more strongly in areas where it is already strongly represented today, i.e. predominantly in the northern German states. This scenario shows a somewhat stronger decrease of 13% to 304 TWh per year.

## Conclusions

If wind energy is expanded to 200 GW of installed capacity in Germany, yield losses of around 10-15% would have to be expected. Nevertheless, it would still generate a lot of energy - more than half of Germany's current electricity demand. This would occur predominantly in winter, at a time when photovoltaics cannot produce much. The losses would be significantly lower than those to be expected from the expansion of offshore wind energy [9]. This simply has to do with the fact that there is much more space available on land and thus the wind turbines can generate energy more efficiently. If they are evenly distributed, then this effect should be smaller, as shown in the 2050A scenario. If the technology develops and the turbines become larger and more powerful - a frequent assumption in scenarios - then these turbines extract more energy from the atmosphere. So the reduction effect should increase accordingly. Overall, this estimate shows that the electricity yield will be lower than assumed in the scenarios mentioned at the beginning [1, 2], so the effect is significant, but not serious.

We can also draw conclusions about the effects on the atmosphere from these calculations. The reduction effect is caused by lower wind speeds, which in these scenarios only decrease to (1 - $\beta$/3 x $f_{cap,max}$ x $Y_{ic}$ )/$Y_{max}$), i.e. by 3.4%. The factor of 1/3 in the equation comes from the fact that wind speeds only enter with the power of 1/3 of the energy extracted (the inverse of the turbine yield, which is proportional to $v^3$) - linearizing then gives the factor of 1/3.

The amount of energy extracted by the wind turbines is really small for the atmosphere: with the expansion of 200 GW, this is about 300 TWh per year of the 12 500 TWh per year dissipated by natural frictional losses. This corresponds to only 2.4%. The large-scale weather patterns of the atmosphere will therefore not feel the wind turbines, especially because the extracted kinetic energy would otherwise have been lost through friction in the boundary layer anyway.

## Summary

*The use of wind energy in Germany is to be expanded to up to 200 Gigawatt by 2050, which is roughly a fourfold increase compared to today. In the process, these wind turbines will extract wind energy, thus affecting the atmosphere. This has an impact on the efficiency of wind energy use because the wind speeds in the affected regions must decrease. This reduction can be*

*estimated using the momentum balance and the associated kinetic energy fluxes, and shows that the electricity yield reduces by about 10 - 15%. The effect is smaller if the wind turbines are more evenly distributed over more space. Despite these effects, a lot of electricity can be generated with wind energy; the scenarios considered would cover more than half of the current electricity demand. However, the effects on the atmosphere are very small. The wind energy generated amounts to only 2.4% of the loss of kinetic energy, which is naturally lost through friction in the lower atmosphere of Germany.*

## Keywords

Wind energy, full load hours, energy transition, kinetic energy, renewable energy, wind, scenarios, climatic impacts

## The author

Axel Kleidon studied physics and meteorology at the University of Hamburg and Purdue University, Indiana, USA. He did his doctoral work at the Max Planck Institute for Meteorology in 1998 on the influence of deep-rooted vegetation on the climate system. He then conducted research at Stanford University in California and at the University of Maryland. Since 2006, he heads an independent research group on "Biospheric Theory and Modelling" at the Max Planck Institute for Biogeochemistry in Jena. His research interests range from the thermodynamics of the Earth system to the natural limits of renewable energy sources.

## Address

Dr Axel Kleidon, Max Planck Institute for Biogeochemistry, Postfach 10 01 64, 07701 Jena. akleidon@bgc-jena.mpg.de

## Literature

[1] https://static.agora-energiewende.de/fileadmin/Projekte/2021/2021_11_DE_KNStrom2035/A-EW_264_KNStrom2035_WEB.pdf
[2] https://www.wind-energie.de/fileadmin/redaktion/dokumente/publikationen-oeffentlich/themen/06-zahlen-und-fakten/20220302_Faktencheck_Wie_viele_Anlagen_braucht_das_Land_final.pdf
[3] https://www.wind-energie.de/fileadmin/redaktion/dokumente/publikationen-oeffentlich/themen/06-zahlen-und-fakten/Factsheet_Status_Windenergieausbau_an_Land_2021.pdf
[4] https://ag-energiebilanzen.de/daten-und-fakten/primaerenergieverbrauch
[5] A. Kleidon, Physik in unserer Zeit **2012**, *43*(3), 136.
[6] A. Kleidon, Meteorol. Z. **2021**, *30*(3), 203.
[7] https://www.ecmwf.int/en/forecasts/datasets/reanalysis-datasets/era5
[8] S. Germer, A. Kleidon, PLoS ONE **2019**, *14*(2), e0211028.
[9] A. Kleidon, Physik in unserer Zeit **2023**, *54*(1), 30.

Supplement: Table 1

|  | Area / km² | Dissipation / W m⁻² | 2021 / MW | 2050A / MW | 2050B / MW |
|---|---|---|---|---|---|
| Baden-Württemberg | 35 751 | 3.3 | 1 730 | 20 008 | 6 164 |
| Bavaria | 70 550 | 3.3 | 2 567 | 39 482 | 9 147 |
| Berlin | 892 | 3.6 | 23 | 499 | 82 |
| Brandenburg | 29 654 | 3.7 | 7 864 | 16 595 | 28 021 |
| Bremen | 419 | 4.3 | 201 | 234 | 716 |
| Hamburg | 755 | 4.3 | 119 | 423 | 424 |
| Hesse | 21 115 | 4.0 | 2 304 | 11 817 | 8 210 |
| Mecklenburg-Vorpommern | 23 213 | 4.0 | 3 567 | 12 991 | 12 710 |
| Lower Saxony | 47 615 | 4.4 | 11 687 | 26 647 | 41 643 |
| North Rhine-Westphalia | 34 110 | 4.3 | 6 388 | 19 089 | 22 761 |
| Rhineland-Palatinate | 19 854 | 4.1 | 3 814 | 11 111 | 13 590 |
| Saarland | 2 570 | 3.9 | 531 | 1 438 | 1 892 |
| Saxony | 18 420 | 4.0 | 1 263 | 10 308 | 4 500 |
| Saxony-Anhalt | 20 452 | 4.0 | 5 318 | 11 446 | 18 949 |
| Schleswig-Holstein | 15 803 | 4.4 | 7 015 | 8 844 | 24 996 |
| Thuringia | 16 202 | 4.3 | 1 739 | 9 067 | 6 196 |
| Total | 357 375 | 3.9 | 56 130 | 200 000 | 200 000 |

*Current installed capacity of wind energy in the different states of Germany as well as in two scenarios for the expansion to 200 GW for the year 2050. The area of the states as well as the average dissipation rate of kinetic energy is also given.*